\documentclass[twocolumn,amsmath,amssymb,nofootinbib,superscriptaddress,showpacs,showkeys]{revtex4}

\usepackage[T1]{fontenc}
\usepackage{ae}
\usepackage{aecompl}
\usepackage[latin1]{inputenc}

\usepackage{graphicx}
\usepackage[hyperfootnotes=false]{hyperref}

\begin{document}


\title{Mathisson's helical motions for a spinning particle --- are they
unphysical?}

\author{L.~Filipe~Costa}
\email{filipezola@fc.up.pt}
\affiliation{Centro de F\'{\i}sica do Porto --- CFP, Departamento de F\'{\i}sica
e Astronomia, Faculdade de Ciências da Universidade do Porto --- FCUP,
Rua do Campo Alegre, 4169-007 Porto, Portugal}

\author{Carlos~Herdeiro}
\email{herdeiro@ua.pt}
\affiliation{Departamento de F\'{\i}sica da Universidade de Aveiro \& I3N, Campus
de Santiago, 3810-183 Aveiro, Portugal}

\author{José~Natário}
\email{jnatar@math.ist.utl.pt}
\affiliation{Departamento de Matemática, Instituto Superior Técnico, 1049-001 Lisboa, Portugal}

\author{Miguel~Zilhão}
\email{mzilhao@fc.up.pt}
\affiliation{Centro de F\'{\i}sica do Porto --- CFP, Departamento de F\'{\i}sica
e Astronomia, Faculdade de Ciências da Universidade do Porto --- FCUP,
Rua do Campo Alegre, 4169-007 Porto, Portugal}

\begin{abstract}
It has been asserted in the literature that Mathisson's helical motions
are unphysical, with the argument that their radius can be arbitrarily
large. We revisit Mathisson's helical motions of a free spinning particle,
and observe that such statement is unfounded. Their radius is finite
and confined to the disk of centroids. We argue that the helical motions
are perfectly valid and physically equivalent descriptions of the
motion of a spinning body, the difference between them being the choice
of the representative point of the particle, thus a \textit{gauge}
choice. We discuss the kinematical explanation of these motions, and
we dynamically interpret them through the concept of hidden momentum.
We also show that, contrary to previous claims, the frequency of the
helical motions coincides, even in the relativistic limit, with the
zitterbewegung frequency of the Dirac equation for the electron. 
\end{abstract}
\maketitle

\section{Introduction}

The equations of motion for spinning pole-dipole particles were first
derived by Mathisson~\cite{MathissonNeueMechanik} in the context
of General Relativity, though similar equations, for the case of \emph{flat}
spacetime, had been derived earlier by Frenkel~\cite{Frenkel} (see
also~\cite{TernovBordovitsyn}) in a special relativistic treatment
applying to a classical model of an electron. These equations have
then been further worked out and re-derived most notably by Weyssenhoff
\cite{WeyssenhoffNature,WeyssenhoffRaabe}, Möller~\cite{MollerAIP},
Bhabha-Corben \cite{BhabhaCorben,Corben,CorbenBook}, Dixon~\cite{Dixon1967}
and Gralla \emph{et al}~\cite{Wald et al}, in the framework of Special
Relativity; and in General Relativity by Papapetrou~\cite{Papapetrou I},
who carried out an exact derivation for pole-dipole particles, Tulczyjew
\cite{Tulczyjew}, Taub~\cite{Taub}, Dixon~\cite{Dixon1964,Dixon1970I}
and Souriau~\cite{Souriau,SouriauCompteRendu}, who made derivations
covariant \emph{at each step}, and more recently Natário~\cite{EulerTop}
and Gralla \emph{et al}~\cite{Wald et al 2010}. To form a determined
system, these equations require a supplementary condition, which amounts
to specify the reference worldline relative to which the moments of
the particle are taken. The natural choice is to require it to be
the center of mass; however by contrast with Newtonian mechanics,
in relativity the center of mass/energy of a spinning particle is
an observer dependent point. Thus, in order to use the concept of
center of mass to fix a worldline of reference, a particular observer
must be specified. This, as shall be explained in detail below, is
done through a spin condition $S^{\alpha\beta}u_{\beta}=0$ (for some
unit time-like vector field $u^{\alpha}$), stating that the reference
worldline is the center of mass as measured by some observer of 4-velocity
$u^{\alpha}$. Its choice can be regarded as a \textit{gauge fixing}.

Mathisson's helical solutions~\cite{Mathisson Zitterbewegung} arise
when one uses the supplementary condition $S^{\alpha\beta}U_{\alpha}=0$,
where $U^{\alpha}$ is the center of mass 4-velocity, thus stating
that the center of mass is measured in its proper frame (i.e., the
frame where it is at rest). This condition was first used by Frenkel
\cite{Frenkel}, and later embodied in the derivation by Mathisson
\cite{MathissonNeueMechanik}, and also employed by Pirani~\cite{Pirani 1956}
as a mean of closing the Papapetrou equations. It will be hereafter
dubbed as the {}``Mathisson-Pirani'' supplementary condition, as
it is best known. The helical motions have been studied since by many
authors (see e.g.~\cite{BhabhaCorben,Corben,CorbenBook,CorbenQRotor,MollerAIP,WeyssenhoffNature,WeyssenhoffRaabe,Semerak II,Plyatsko Non-Oscillatory}).

These helical motions exist even for a free particle in flat spacetime,
and are still rather mysterious today. They were first~\cite{Mathisson Zitterbewegung}
interpreted, for the case of the electron, as the classical counterpart
of the `zitterbewegung' observed in Dirac's equation, based on the
coincidence of frequencies obtained in the non-relativistic limit;
this point of view was then supported by other authors, e.g.~\cite{Corben,CorbenBook,Papapetrou I,CorbenQRotor}.
Möller~\cite{MollerAIP} provided a kinematical interpretation of
the helices as arising from the motions of what he called the {}``pseudo-centers
of mass'' (see equivalent treatment in Sec.~\ref{sec:Kinematical-explanation-of}
below). In spite of this, they have been deemed unphysical~\cite{Dixon1964,Dixon1965}
with the argument that the radius of the helices can be arbitrarily
large~\cite{WeyssenhoffNature,WeyssenhoffRaabe,Dixon1964,Dixon1965,DixonReview},
which would be contradicted by experiment (this was actually what
initially motivated Dixon's multipole approach to extended bodies
\cite{Dixon1964}, embodying the alternative condition $S^{\alpha\beta}P_{\beta}=0$,
proposed by Tulczyjew). This would make them also inconsistent with
Möller's scheme. Also, it was argued~\cite{WeyssenhoffNature,WeyssenhoffRaabe}
that the coincidence with the frequency of Dirac's equation zitterbewegung
motions holds only in the non-relativistic limit.

Herein we will show that the assessments regarding the unphysical
nature of the helical motions are unfounded and originate from a mistake
in the treatment in~\cite{WeyssenhoffNature,WeyssenhoffRaabe}. We
argue otherwise: that they are physically acceptable, being actually
alternative and equivalent (albeit more complicated) descriptions
of the motion of a spinning body; these different descriptions are
a matter of choice, resulting from the incompleteness of the gauge
fixing provided by the Mathisson-Pirani supplementary condition, which
leaves a \textit{residual gauge freedom}. Their radius is shown to
be contained within the disk of centroids, whose size is actually
the \emph{minimum} size a classical spinning particle can have if
it is to have finite angular momentum and positive mass without violating
Special Relativity. We kinematically explain these solutions, showing
they are consistent with Möller's interpretation, and we show they
are dynamically consistent descriptions of the motion of the body,
which can be understood through the concept of {}``hidden momentum''.

Finally, regarding the correspondence with the quantum problem, we
point out that the assertions in~\cite{WeyssenhoffNature,WeyssenhoffRaabe}
that the frequencies only coincide in the non-relativistic limit originate
from the same mistake that leads to the arbitrarily large radius;
indeed the frequencies coincide exactly.

\subsection{Notation and conventions\label{sub:Notation-and-conventions}}

We summarize here the conventions that we use for the various quantities
necessary to describe the motion of a spinning particle:

\begin{enumerate}
\item $U^{\alpha}\equiv dz^{\alpha}/d\tau$ is the tangent vector to the
worldline of reference $z^{\alpha}(\tau)$ (in this work it amounts
to the 4-velocity of some suitably defined center of mass); 
\item $u^{\alpha}$ denotes a generic unit time-like vector defined along
the worldline of reference; it can be thought as the instantaneous
4-velocity of an observer $\mathcal{O}(u)$; 
\item $\Sigma(z,u)$ is the hypersurface generated by all geodesics orthogonal
to $u^{\alpha}$ at a point $z^{\alpha}$; in flat spacetime, it is
simply the 3-space orthogonal to $u^{\alpha}$ (it can be thought
of as the instantaneous rest space of $\mathcal{O}(u)$); 
\item $x_{CM}^{\alpha}(u)$ is the center of mass as measured in the instantaneous
rest space of $\mathcal{O}(u)$; 
\item \emph{Centroids}: following~\cite{Gravitation}, we dub the centers
of mass $x_{CM}^{\alpha}(u)$ as measured by arbitrary observers as
centroids; these divide in two subclasses: 1) \emph{proper center
of mass} $x_{CM}(U)$ --- center of mass as measured in its own rest
frame; 2) \emph{non-proper center of mass} --- center of mass measured
by an observer not comoving with it. Sometimes we shall use the abbreviation
CM for center of mass. 
\item \emph{Masses:} $m(u)\equiv-P^{\alpha}u_{\alpha}$ denotes the mass
as measured by $\mathcal{O}(u)$; by $m\equiv m(U)=-P^{\alpha}U_{\alpha}$
we denote the proper mass (i.e., the mass measured in the CM frame);
and $M\equiv\sqrt{-P^{\alpha}P_{\alpha}}$ is the mass as measured
in the zero 3-momentum frame. 
\item \label{enu:Sstar}We denote by $S_{\star}^{\alpha\beta}$ the angular
momentum tensor about the centroid $x_{CM}(P)$ measured in the zero
3-momentum frame (i.e., $S_{\star}^{\alpha\beta}P_{\beta}=0$), and
by $S_{\star}^{\alpha}$ the corresponding spin vector, obeying $S_{\star}^{\alpha\beta}=\epsilon_{\ \ \mu\nu}^{\alpha\beta}S_{\star}^{\mu}P^{\nu}/M$. 
\item $\epsilon_{\alpha\beta\gamma\delta}$ denotes the Levi-Civita tensor;
we choose $\epsilon_{0123}=-1$ (for flat spacetime). We denote by
$\vec{A}\times_{U}\!\vec{B}$ the spatial part of the vector $\epsilon_{\ \beta\gamma\delta}^{\alpha}A^{\beta}B^{\gamma}U^{\delta}$
with respect to a given frame $\mathcal{O}(u)$; and $\vec{A}\times\vec{B}\equiv\vec{A}\times_{u}\!\vec{B}$. 
\end{enumerate}

\section{Equations of motion for free spinning particles in flat spacetime}

In a multipole expansion, a body is represented by a set of moments
of $T^{\alpha\beta}$, called {}``inertial'' or {}``gravitational''
moments (forming the so called~\cite{MathissonNeueMechanik} {}``gravitational
skeleton''). The moments are taken about a reference worldline $z^{\alpha}(\tau)$,
which will be chosen as a suitably defined center of mass to be discussed
below. Truncating the expansion at dipole order, the equations of
motion involve only two moments of $T^{\alpha\beta}$~\cite{DixonReview,Dixon1964,Dixon1967},
the momentum $P^{\alpha}$ and the angular momentum $S^{\alpha\beta}$:
\begin{eqnarray}
P^{\alpha} & \equiv & \int_{\Sigma(\tau,u)}T^{\alpha\beta}d\Sigma_{\beta}\ ,\label{eq:Pgeneral}\\
S^{\alpha\beta} & \equiv & 2\int_{\Sigma(\tau,u)}r^{[\alpha}T^{\beta]\gamma}d\Sigma_{\gamma}\ .\label{eq:Sab}\end{eqnarray}
 Here $P^{\alpha}(\tau)$ is the 4-momentum of the body; $S^{\alpha\beta}(\tau)$
is the angular momentum about a point $z^{\alpha}(\tau)$ of the reference
worldline; $\Sigma(\tau,u)\equiv\Sigma(z(\tau),u)$; $r^{\alpha}\equiv x^{\alpha}-z^{\alpha}(\tau)$,
where $\{x^{\alpha}\}$ is a chart on spacetime; and finally $d\Sigma_{\gamma}\equiv n_{\gamma}d\Sigma$,
where $n_{\gamma}$ is the (past-pointing) unit normal to $\Sigma(\tau,u)$
and $d\Sigma$ is the $3$-volume element on $\Sigma(\tau,u)$.

For simplicity, we will consider the background to be Minkowski spacetime
without any further fields. In this case $P^{\alpha},\ S^{\alpha\beta}$
are independent of $\Sigma$, and the equations of motion that follow
from the conservation law $T_{\ \ ;\beta}^{\alpha\beta}=0$ are~\cite{WeyssenhoffRaabe,Papapetrou I,Dixon1964,Dixon1967,Dixon1970I,DixonReview}:
\begin{equation}
\frac{DP^{\alpha}}{d\tau}=0\quad(a),\qquad\frac{DS^{\alpha\beta}}{d\tau}=2P^{[\alpha}U^{\beta]}\quad(b).\label{eq:Eqs Motion Flat}\end{equation}
 Contracting (\ref{eq:Eqs Motion Flat}b) with $U^{\alpha}$ we obtain
an expression for the momentum \begin{equation}
P^{\alpha}=mU^{\alpha}-\frac{DS^{\alpha\beta}}{d\tau}U_{\beta}\ ,\label{eq:Momentum general}\end{equation}
 where $m\equiv-P^{\alpha}U_{\alpha}$. Eqs.~\eqref{eq:Eqs Motion Flat}
form an indeterminate system. Indeed, there are $13$ unknowns ($P^{\alpha}$,
$3$ independent components of $U^{\alpha}$, and $6$ independent
components of $S^{\alpha\beta}$) for only $10$ equations%
\footnote{Substituting (\ref{eq:Momentum general}) in (\ref{eq:Eqs Motion Flat}),
we obtain the Eqs. in Mathisson's representation~\cite{MathissonNeueMechanik,Papapetrou I,Pirani 1956,DixonReview};
in this case we would have 10 independent unknowns ($m$, 3 independent
components of $U^{\alpha}$, and 6 of $S^{\alpha\beta}$), for 7 independent
equations: 4 from (\ref{eq:Eqs Motion Flat}a) and only 3 from (\ref{eq:Eqs Motion Flat}b),
since contracting the latter with $U^{\alpha}$ leads to an identity.%
}. To close the system we need to specify the representative point
of the body (i.e., the worldline of reference relative to which $S^{\alpha\beta}$
is taken). That can be done through a supplementary spin condition
of the type $S^{\alpha\beta}u_{\beta}=0$, where $u^{\alpha}(\tau)$
is some appropriately chosen unit timelike vector, which effectively
kills off $3$ components of the angular momentum; this condition,
as we shall see in the next section, means that the reference worldline
is the center of mass as measured in the rest frame of the observer
of velocity $u^{\alpha}$. Hence $U^{\alpha}$ is the center of mass
4-velocity and $m$ denotes the \emph{proper mass}, i.e., the energy
of the body as measured in the center of mass frame.


We note from Eq.~\eqref{eq:Momentum general} that the momentum of
a spinning particle is not, in general, parallel to its 4-velocity;
it is said to possess {}``hidden momentum''~\cite{Wald et al 2010,Gyros},
which will play a key role in this discussion.

\section{Center of mass and the significance of the spin supplementary condition\label{sub:Center-of-mass}}

\begin{figure}
\includegraphics[width=0.45\textwidth]{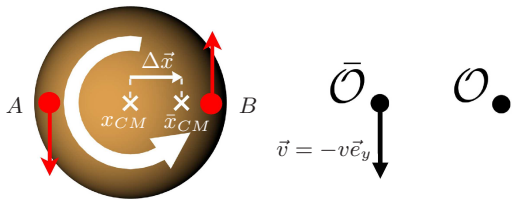}

\caption{\label{fig:CMShift} Center of mass of a spinning particle ($\vec{S}=S\vec{e}_{z}$,
orthogonal to the page) as evaluated by two different observers. Observer
$\mathcal{O}$ is at rest with respect to center of mass $x_{CM}^{i}\equiv x_{CM}^{i}(u)$
it measures (i.e., $x_{CM}^{i}$ is a proper center of mass). Observer
$\bar{\mathcal{O}}$, moving with velocity $\vec{v}=-v\vec{e}_{y}$
relative to $\mathcal{O}$, sees the points on the right hemisphere
(e.g. point $B$) moving faster than the points in the left hemisphere
(e.g. point $A$), and, therefore, for $\bar{\mathcal{O}}$, the right
hemisphere will be more massive than the left one. This means that
the center of mass $\bar{x}_{CM}^{i}\equiv x_{CM}^{i}(\bar{u})$ as
evaluated in the moving frame of $\bar{\mathcal{O}}$ is shifted to
the right (relative to $x_{CM}$). If $u^{\alpha}=P^{\alpha}/M$,
the shift is exactly $\Delta\vec{x}=\vec{S}_{\star}\times\vec{v}/M$.}

\end{figure}

In relativistic physics, the center of mass of a spinning particle
is observer dependent. This is illustrated in Fig.~\ref{fig:CMShift}.
%

Thus one needs to specify the frame in which the center of mass is
to be evaluated. That can be done through a spin condition of the
type $S^{\alpha\beta}u_{\beta}=0$, as we will show next. The vector
$(d_{G}^{u})^{\alpha}\equiv-S^{\alpha\beta}u_{\beta}$ yields the
{}``mass dipole moment'' as measured in the rest frame of the observer
$\mathcal{O}$ of 4-velocity $u^{\alpha}$. This is easily seen in
this frame, where $u^{i}=0$ and $S^{\alpha\beta}u_{\beta}=S^{\alpha0}u_{0}$.
Thus, from Eq.~\eqref{eq:Sab}: \begin{equation}
S^{i0}=2\int_{\Sigma(\tau,u)}r^{[i}T^{0]\gamma}d\Sigma_{\gamma}=\int x^{i}T^{00}d^{3}x-m(u)z^{i}\ ,\label{eq:Massdipole}\end{equation}
 where, as before, $r^{\alpha}\equiv x^{\alpha}-z^{\alpha}$ (note
that $r^{0}=0$, since the integration is performed in the hypersurface
$\Sigma(u)$ orthogonal to $u^{\alpha}$), and $m(u)\equiv-P^{\alpha}u_{\alpha}$
denotes the mass as measured in the frame $\mathcal{O}$. The first
term of (\ref{eq:Massdipole}) is by definition $m(u)x_{CM}^{i}(u)$,
where $x_{CM}^{i}(u)$ are the coordinates of the center of mass as
measured by $\mathcal{O}$, and so \begin{equation}
x_{CM}^{i}(u)-z^{i}=\frac{S^{i0}}{m(u)}\Leftrightarrow x_{CM}^{\alpha}(u)-z^{\alpha}=-\frac{S^{\alpha\beta}u_{\beta}}{m(u)}\ .\label{eq:CMCovariantShift}\end{equation}
 Thus we see that the condition $S^{\alpha\beta}u_{\beta}=0$ is precisely
the condition that the reference worldline $z^{\alpha}(\tau)$ is
the center of mass as measured in this frame. It allows us to write
$S^{\alpha\beta}=\epsilon_{\ \ \ \gamma\delta}^{\alpha\beta}S^{\gamma}u^{\delta}$,
where $S^{\alpha}(u)$ is the spin 4-vector, defined as being the
4-vector with components $(0,\vec{S})$ in the rest frame of $\mathcal{O}$,
that is, $S^{\alpha}=\frac{1}{2}\epsilon_{\ \ \beta\gamma\delta}^{\alpha}u^{\beta}S^{\gamma\delta}$.
In order to see how the center of mass position changes in a change
of observer, consider now another observer $\bar{\mathcal{O}}$ moving
relative to $\mathcal{O}$ with 4-velocity $\bar{u}^{\alpha}=\bar{u}^{0}(1,\vec{v})$;
for this observer the center of mass will be at a different position,
as depicted in Fig.~\ref{fig:CMShift}. The {}``mass dipole moment''
as measured by $\bar{\mathcal{O}}$ is%
\footnote{\label{fn:Hypersurface}We have shown in discussion above and in Eq.
(\ref{eq:Massdipole}) that $(d_{G}^{u})^{\alpha}\equiv-S^{\alpha\beta}u_{\beta}$,
with $S^{\alpha\beta}$ defined integrating in an hypersurface $\Sigma(\tau,u)$
orthogonal to $u^{\alpha}$, yields the mass dipole as measured by
$\mathcal{O}$. Note that it follows from the conservation laws $T_{\ \ ;\beta}^{\alpha\beta}=0$
that the \emph{2-tensor} $\mathbf{S}$ does not depend on $\Sigma$.
Only its \emph{components} $S^{\alpha\beta}$ do, since a choice of
$\Sigma$ amounts in this case to \emph{choose the frame} where $S^{\alpha\beta}$
are expressed (see e.g.~\cite{Gravitation}). Hence $(d_{G}^{\bar{u}})^{\alpha}=-S^{\alpha\beta}\bar{u}_{\beta}$,
with $S^{\alpha\beta}$ again defined with respect to $\Sigma(\tau,u)$,
yields indeed the mass dipole measured by $\bar{\mathcal{O}}$, only
written in the coordinates of $\mathcal{O}$. But since $\vec{d}_{G}(\bar{u})\perp\vec{v}$,
cf. Eq. (\ref{eq:CMShiftGeneral}), the coordinates $(d_{G}^{\bar{u}})^{\alpha}$
are actually the same in the systems of $\mathcal{O}$ and $\bar{\mathcal{O}}$.%
} $(d_{G}^{\bar{u}})^{\alpha}=-S^{\alpha\beta}\bar{u}_{\beta}$; thus,
the center of mass as measured by $\bar{\mathcal{O}}$ is displaced
by a vector $\Delta x^{\alpha}=-S^{\alpha\beta}\bar{u}_{\beta}/m(\bar{u})$
relative to the reference worldline $z^{\alpha}$, where $m(\bar{u})\equiv-P^{\gamma}\bar{u}_{\gamma}$
denotes the mass of the particle as measured by $\bar{\mathcal{O}}$.
Hence we get: \begin{eqnarray}
\Delta x^{i} & = & \frac{1}{P^{\gamma}\bar{u}_{\gamma}}\left(S^{i0}\bar{u}_{0}+S^{ij}\bar{u}_{j}\right)\ =\ \frac{(\vec{S}\times\vec{v})^{i}}{P^{0}-P^{i}v_{i}}\ ,\label{eq:CMShiftGeneral}\end{eqnarray}
 (recall that, in this frame, $S^{i0}=x_{CM}^{i}(u)-z^{i}=0$, since
we chose $x_{CM}^{i}(u)$ as the reference worldline). Note that the
coordinates of the 3-vector $\Delta x^{i}$ are the same in the frame
$\mathcal{O}$ or $\bar{\mathcal{O}}$, since $\Delta\vec{x}\perp\vec{v}$.
If $u^{\alpha}=P^{\alpha}/M$, $M\equiv\sqrt{-P^{\alpha}P_{\alpha}}$,
i.e., if we take as reference worldline the center of mass $x_{CM}(P)$
as measured in the zero 3-momentum frame, then:\begin{eqnarray}
\Delta x^{i} & = & \ \frac{(\vec{S}_{\star}\times\vec{v})^{i}}{M}\ ,\label{eq:CMShift}\end{eqnarray}
 where we denote by $S_{\star}^{\alpha\beta}$ the angular momentum
with respect to $x_{CM}(P)$, cf. point \ref{enu:Sstar} of Sec. \ref{sub:Notation-and-conventions}.
In general one wants the equations of motion not to depend on quantities
(the center of mass) measured by a particular observer, but instead
a center of mass defined only in terms of properties {}``intrinsic''
to the particle. Two conditions accomplishing this are frequently
found in the literature%
\footnote{A review (with a comprehensive list of references) on the literature
regarding this subject may be found in~\cite{Semerak I}.%
}: the Mathisson-Pirani condition $S^{\alpha\beta}U_{\beta}=0$ (that
is, $u^{\alpha}=U^{\alpha}$) and the Tulczyjew-Dixon condition $S^{\alpha\beta}P_{\beta}=0$
(that is, $u^{\alpha}=P^{\alpha}/M$). The latter amounts to take
as reference worldline the center of mass as measured in the frame
of zero 3-momentum, $P^{i}=0$; the former comes as the most natural
choice, as it amounts to compute the center of mass in its proper
frame, i.e., \emph{in the frame where it has zero 3-velocity}. Such
center of mass is dubbed a {}``proper center of mass''~\cite{MollerAIP}.

\section{Mathisson's Helical Solutions}

Using the Mathisson-Pirani condition $S^{\alpha\beta}U_{\beta}=0$,
implying $S^{\alpha\beta}=\epsilon^{\alpha\beta\mu\nu}S_{\mu}U_{\nu},$
we can rewrite \eqref{eq:Momentum general} as \begin{equation}
P^{\alpha}=mU^{\alpha}+S^{\alpha\beta}a_{\beta}=mU^{\alpha}+\epsilon_{\ \ \gamma\delta}^{\alpha\beta}a_{\beta}S^{\gamma}U^{\delta}\ ,\label{eq:momentumhelical}\end{equation}
 where $a^{\alpha}=DU^{\alpha}/d\tau$. It follows from Eqs.~(\ref{eq:Eqs Motion Flat}a)
and (\ref{eq:momentumhelical}) that the proper mass $m=-P^{\alpha}U_{\alpha}$
is a \emph{constant} of the motion: $dm/d\tau=0$. Eq. (\ref{eq:Eqs Motion Flat}b)
can be written as $DS^{\mu}/d\tau=a_{\nu}S^{\nu}U^{\mu}$, stating
that the spin vector $S^{\alpha}$ is Fermi-Walker transported along
the CM worldline. This equation, coupled with (\ref{eq:momentumhelical})
and (\ref{eq:Eqs Motion Flat}a), effectively means that the spin
vector is parallel transported:\[
\frac{DS^{\alpha}}{d\tau}=0\ ,\]
 since, as can be seen substituting (\ref{eq:momentumhelical}) in
(\ref{eq:Eqs Motion Flat}a) and contracting with $S^{\alpha}$, the
spin vector is orthogonal to the acceleration: $a_{\nu}S^{\nu}=0$.
Noting, from Eq. (\ref{eq:momentumhelical}), that $P^{\alpha}S_{\alpha}=0$,
we can take, without loss of generality, the constant spin vector
pointing along the $z$-axis,\[
S^{\alpha}=(0,0,0,S)\ ,\]
 in the global Cartesian frame of zero 3-momentum, \[
P^{\alpha}=(M,0,0,0)=(\chi m,0,0,0)\ .\]
 Here $M\equiv\sqrt{P^{\alpha}P_{\alpha}}$ denotes the mass/energy
of the particle as measured in this frame, and $\chi$ is some constant
to be determined in the course of the derivation. The equations of
motion to be solved are \eqref{eq:momentumhelical}. These require
$U^{t}=1/\chi$, $U^{z}=0$ and \[
U^{x}+\frac{S}{m\chi}\frac{dU^{y}}{d\tau}=0\ ,\qquad U^{y}-\frac{S}{m\chi}\frac{dU^{x}}{d\tau}=0\ .\]
 The general solution for the worldline of reference describes the
famous \textit{helical motions}, which correspond to \textit{clockwise}
(i.e.~\emph{opposite} to the spin direction) circular motions with
radius $R$ and speed $v$ on the $xy$ plane, \begin{equation}
z^{\alpha}(\tau)=\left(\gamma\tau,-R\cos\left(\frac{v\gamma}{R}\tau\right),R\sin\left(\frac{v\gamma}{R}\tau\right),0\right),\label{eq:PositionMathisson}\end{equation}
 (where $\tau$ is the proper time and $\gamma=-P^{\alpha}U_{\alpha}/M=1/\sqrt{1-v^{2}}=1/\chi$).
With these choices the 4-velocity takes the form \begin{equation}
U^{\alpha}=\left(\gamma,v\gamma\sin\left(\frac{v\gamma}{R}\tau\right),v\gamma\cos\left(\frac{v\gamma}{R}\tau\right),0\right),\label{eq:4VelMathisson}\end{equation}
 and the acceleration is\[
a^{\alpha}=\frac{v^{2}\gamma^{2}}{R}\left(0,\cos\left(\frac{v\gamma}{R}\tau\right),-\sin\left(\frac{v\gamma}{R}\tau\right),0\right)\ .\]
 The equations of motion require that the radius of the trajectory
obeys \begin{equation}
R=\frac{v\gamma^{2}S}{m}\ .\label{eq:RadiusWeyssenhoff}\end{equation}

All these helical solutions are equivalent descriptions of the motion
of a spinning body, the difference between them being the representative
point they use to describe the body. Note also that (this is true
in flat spacetime, and in the absence of electromagnetic field) the
non-helical solution $R=0$ corresponds to $P^{\alpha}\parallel U^{\alpha}$,
i.e., to the (unique) solution defined by the Tulczyjew-Dixon condition
$S^{\alpha\beta}P_{\beta}=0$.

The fact that $\gamma$ in Eq.~\eqref{eq:RadiusWeyssenhoff} can
be arbitrarily large has led some authors~\cite{WeyssenhoffNature,WeyssenhoffRaabe,Dixon1964,Dixon1965,DixonReview}
to believe that the same extended body may be represented by circular
trajectories with any radius. That would be inconsistent with the
results in Sec.~\ref{sec:Kinematical-explanation-of} below, with
Möller's treatment in ref.~\cite{MollerAIP}, and with the results
in e.g.~\cite{Semerak II}. This is not the case, however; as we
shall now show, indeed the radius is finite (and confined to the disk
of centroids, cf. Fig. \ref{fig:MolerPicture}), and the misunderstanding
originates from the fact that keeping the parameters $m$ and $S$
fixed {\em does not} correspond to considering the same extended
body.

In a multipole expansion, an extended body is characterized by its
multipole moments. In the pole-dipole approximation, that amounts
to specify its momentum $P^{\alpha}$ and its spin tensor $S^{\alpha\beta}$.
These, cf. Eqs.~\eqref{eq:Pgeneral}-\eqref{eq:Sab}, are defined
with respect to an hypersurface of integration $\Sigma$ (which is
interpreted as the rest space of the observer, see also Footnote \ref{fn:Hypersurface}),
and, in the case of $S^{\alpha\beta}$, also with respect to a reference
worldline $z^{\alpha}$. Different representations of the same extended
body must yield the same moments with respect to \emph{the same observer}
and the \emph{same reference worldline}. So instead of $m=-U^{\alpha}P_{\alpha}$
(which depends, via $U^{\alpha}$, on the particular helix chosen),
we must in fact fix \begin{equation}
M=\sqrt{-P_{\alpha}P^{\alpha}}=\frac{m}{\gamma}\ .\end{equation}

Similarly, it is \emph{not} the spin vector $S^{\alpha}$ (nor the
spin tensor $S^{\alpha\beta}$ obeying $S^{\alpha\beta}U_{\beta}=0$)
that we must fix for different trajectories representing the same
extended body. By choosing the Mathisson-Pirani condition, the tensor
$S^{\alpha\beta}$ showing up in the equations (\ref{eq:Eqs Motion Flat})
is always orthogonal to $U^{\alpha}$; as explained in section \ref{sub:Center-of-mass},
that means that $S^{\alpha\beta}$ is the angular momentum evaluated
with respect to $x_{CM}^{\alpha}(U)$, i.e., the center of mass as
measured by the observer of 4-velocity $U^{\alpha}$. Let $U^{\alpha}$
and $\bar{U}^{\alpha}$ denote the 4-velocity vectors, Eq.~\eqref{eq:4VelMathisson},
of two different helical representations. The tensor $S^{\alpha\beta}$,
obeying $S^{\alpha\beta}U_{\beta}=0$, must be, in general, different
from the tensor $\bar{S}^{\alpha\beta}$, obeying $\bar{S}^{\alpha\beta}\bar{U}_{\beta}=0$,
\emph{if} $S^{\alpha\beta}$ \emph{and} $\bar{S}^{\alpha\beta}$ \emph{are
to represent the same body}, since the former is the angular momentum
about the point $x_{CM}^{\alpha}(U)$, and the latter about the point
$x_{CM}^{\alpha}(\bar{U})$.

Let $S_{\star}^{\alpha\beta}$ denote the spin tensor for the \emph{non-helical}
trajectory (corresponding to $R=0$, $\tau=t$, $z^{\alpha}(\tau)=\tau\delta_{0}^{\alpha}$
and $U^{\alpha}=P^{\alpha}/M$), \begin{equation}
S_{\star}^{\alpha\beta}=2\int_{\Sigma(\tau,P)}r^{[\alpha}T^{\beta]\gamma}d\Sigma_{\gamma}\ \end{equation}
 with $r^{\alpha}=x^{\alpha}-z^{\alpha}(\tau)$. This corresponds
to a spin vector \begin{equation}
S_{\star}^{\alpha}=(0,0,0,S_{\star})\ ,\end{equation}
 and so \begin{equation}
S_{\star}^{\alpha\beta}=\epsilon_{\,\,\,\,\,\,\mu\nu}^{\alpha\beta}S_{\star}^{\mu}U^{\nu}=\epsilon_{\,\,\,\,\,\,30}^{\alpha\beta}S_{\star}=\epsilon^{0\alpha\beta3}S_{\star}\ .\end{equation}
 Therefore the non-vanishing components of $S_{\star}^{\alpha\beta}$
are $S_{\star}^{12}=-S_{\star}^{21}=S_{\star}$. The spin tensor for
a helical trajectory, however, is (in the same global Cartesian frame
of zero 3-momentum%
\footnote{Note that integrating in the hypersurface $\Sigma(\tau,P)$, orthogonal
to $P^{\alpha}$, amounts to write $S^{\alpha\beta}$ in the frame
$P^{i}=0$, see Footnote \ref{fn:Hypersurface}.%
}): \begin{equation}
S^{\alpha\beta}=2\int_{\Sigma(\tau,P)}\bar{r}^{[\alpha}T^{\beta]\gamma}d\Sigma_{\gamma}\ ,\end{equation}
 where $\bar{r}^{\alpha}=x^{\alpha}-\bar{z}^{\alpha}(\tau)$, $\bar{z}^{\alpha}(\tau)$
being given by Eq.~\eqref{eq:PositionMathisson} with $R\ne0$. Therefore
\begin{equation}
S^{\alpha\beta}=S_{\star}^{\alpha\beta}-2\bar{z}^{[\alpha}(\tau)P^{\beta]}\ .\label{eq:SpinTensorRelation}\end{equation}
 The condition $S^{\alpha\beta}U_{\beta}=0$ yields \begin{equation}
S_{\star}^{\alpha\beta}U_{\beta}+m\bar{z}^{\alpha}(\tau)-\gamma^{2}\tau P^{\alpha}=0\ ,\end{equation}
 which when written out in components reduces to \begin{equation}
mR=v\gamma S_{\star}\ ,\end{equation}
 and so \begin{equation}
S_{\star}=\gamma S\ .\end{equation}

The fixed quantities for a given body are then $\gamma S=S_{\star}$
and $m/\gamma=M$, not $m$ and $S$. Accordingly, the same extended
body will be represented by helical trajectories whose radius $R$
satisfies \begin{equation}
R=\frac{vS_{\star}}{M}\ ,\label{eq:Radius Fixed}\end{equation}
 and thus must be smaller than $S_{\star}/M$. The angular frequency
of the helices is, from Eqs.~\eqref{eq:PositionMathisson} and \eqref{eq:Radius Fixed}:
\begin{equation}
\omega=\frac{v}{R}=\frac{M}{S_{\star}}=\frac{m}{\gamma^{2}S}\ ,\label{eq:wMathisson}\end{equation}
 which is thus the same for all helical solutions representing the
same extended body. As we shall see in the next section, this is entirely
consistent with Möller's picture of the disk, rotating \emph{rigidly}
with frequency $\omega$, formed by the many proper centers of mass.

\section{Kinematical explanation of the helical motions\label{sec:Kinematical-explanation-of}}

In this section we will provide a kinematical explanation for the
helical motions. Although stated and derived in a different form,
it is equivalent to Möller's treatment in~\cite{MollerAIP}, which
does not seem to be well understood in the literature.

\begin{figure}
\includegraphics[scale=0.6]{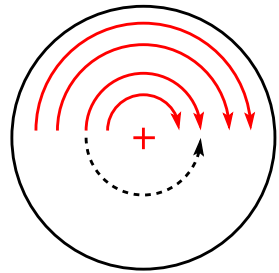}

\caption{\label{fig:MolerPicture}Kinematical explanation of the helical motions
allowed by $S^{\alpha\beta}U_{\beta}=0$: every point within a disk
of radius $S_{\star}/M$ is a \emph{centroid} corresponding to some
observer; and it is also a \emph{proper} center of mass \emph{if it
rotates with angular velocity} $\omega=M/S_{\star}$ \emph{in the
opposite sense of the spinning body} (solid red lines).}

\end{figure}

The origin of the helical motions is the fact that the condition $S^{\alpha\beta}U_{\beta}=0$
does not determine the reference worldline uniquely (i.e., it \textit{does
not fix completely the gauge freedom}). In other words, there is not
a unique answer to the question: which is the center of mass such
that it is at rest relative to the frame where it is evaluated? In
order to see this, consider for simplicity a free particle in flat
spacetime. Clearly, for this case, \emph{one of the solutions} of
Eqs.~\eqref{eq:Eqs Motion Flat} supplemented by $S^{\alpha\beta}U_{\beta}=0$
is straight line motion, with $U^{\alpha}=P^{\alpha}/M$ constant.
Let $\mathcal{O}$ be the observer of 4-velocity $u^{\alpha}=P^{\alpha}/M$
(i.e., its rest frame is the zero 3-momentum frame). The center of
mass as measured by this observer is the point $x_{CM}^{\alpha}(P)$
in Fig.~\ref{fig:CMShift}. This point is at rest relative to $\mathcal{O}$,
so that it is clearly a proper center of mass. But now let again $\bar{\mathcal{O}}$
be an observer moving relative to $\mathcal{O}$ with 3-velocity $\vec{v}$.
The 4-velocity of $\bar{\mathcal{O}}$ is $\bar{u}^{\alpha}=\gamma(u^{\alpha}+v^{\alpha})$,
where $\gamma\equiv-u_{\alpha}\bar{u}^{\alpha}$ and $v^{\alpha}$
is the relative velocity vector which is spatial with respect to $u^{\alpha}$.
Observer $\bar{\mathcal{O}}$ measures the center of mass $x_{CM}^{\alpha}(\bar{u})$
(i.e., its centroid) to be at a different position, as shown by Eq.~\eqref{eq:CMShift};
and in general that point will \emph{not} be a proper center of mass,
since it will be moving relative to $\bar{\mathcal{O}}$. Choosing
$x_{CM}^{\alpha}(P)$ as our reference worldline ($z^{\alpha}=x_{CM}^{\alpha}(P)$),
let $\tau_{P}$ be the proper time along it. The relative position
$\Delta x^{\alpha}=x_{CM}^{\alpha}(\bar{u})-x_{CM}^{\alpha}(P)$ is
the spatial (with respect to $u^{\alpha}$) vector $\Delta x^{\alpha}=S_{\star}^{\alpha\beta}\bar{u}_{\beta}/P^{\gamma}u_{\gamma}=-S_{\star}^{\alpha\beta}v_{\beta}/M$.
Noting, from Eqs.~\eqref{eq:Eqs Motion Flat}, that $DS_{\star}^{\alpha\beta}/d\tau_{P}=0$,
it evolves along $z^{\alpha}(\tau_{P})$ as\begin{equation}
\frac{D\Delta x^{\alpha}}{d\tau_{P}}=-\frac{S_{\star}^{\alpha\beta}}{M}\frac{Dv_{\beta}}{d\tau_{P}}\ \;(a)\quad\Leftrightarrow\ \frac{d\vec{\Delta x}}{dt}=\frac{\vec{S}_{\star}\times\vec{a}_{c}}{M}\ \;(b).\label{eq:ShiftVel}\end{equation}
 The second equation holds in the rest frame of $\mathcal{O}$ (the
frame $u^{i}=0=P^{i}$), where the time coordinate is $t=\tau_{P}$,
and $\vec{a}_{c}\equiv d\vec{v}/dt$ denotes the \emph{coordinate}
acceleration of observer $\bar{\mathcal{O}}$ in the frame of $\mathcal{O}$.
Note that it can be directly obtained from (\ref{eq:CMShift}) by
simply differentiating with respect to the coordinate $t$. Thus we
see that if $v^{\alpha}$ is parallel transported along $z^{\alpha}(\tau_{P})$,
which in flat spacetime is ensured by taking $\bar{\mathcal{O}}$
inertial, then $D\Delta x^{\alpha}/d\tau_{P}=0$, implying that $x_{CM}^{\alpha}(\bar{u})$
is fixed relative to $x_{CM}^{\alpha}(P)$; and thus moves relative
to $\bar{\mathcal{O}}$ at a speed $-\vec{v}$. The set of centroids
measured by all the possible \emph{inertial} observers forms a disk
of points \emph{all at rest} with respect to each other and (again,
for a \emph{free particle in flat spacetime}) with respect to $x_{CM}^{\alpha}(P)$,
around which the disk is centered. However if we consider $\bar{\mathcal{O}}$
to be accelerating, then the velocity $\vec{v}_{CM}(\bar{u})=d\Delta\vec{x}/dt$
of the centroid he measures changes in a non-trivial way, as shown
by Eq.~\eqref{eq:ShiftVel}. Now if we take the case that $\bar{\mathcal{O}}$
itself also moves with 3-velocity\begin{equation}
\vec{v}=\frac{\vec{S}_{\star}\times\vec{a}_{c}}{M}\equiv\frac{1}{M}\left(\vec{S}_{\star}\times\frac{d\vec{v}}{dt}\right)\ ,\label{eq:VelHelical}\end{equation}
 then $x_{CM}^{\alpha}(\bar{u})$ is at rest relative to $\bar{\mathcal{O}}$,
i.e., it is a \emph{proper} center of mass. The solutions of Eq.~\eqref{eq:VelHelical}
are circular motions in the plane orthogonal to $\vec{S}_{\star}$,
with radius $R=\Delta x=|\vec{v}\times\vec{S}_{\star}|/M$, and angular
velocity $\vec{\omega}=-M\vec{S}_{\star}/{S_{\star}}^{2}$. Note that
the angular velocity is \emph{constant} (does not depend on $R$)
and is in \emph{opposite sense} to the rotation of the body. Hence
the set of all possible \emph{proper} centers of mass fills a disk
of radius $\Delta x_{max}=S_{\star}/M$ (i.e., of the same size of
the disk of centroids) in the plane orthogonal to $\vec{S}_{\star}$,
\emph{counter-rotating rigidly} with angular velocity $\vec{\omega}$.
In other words: from Eq.~\eqref{eq:CMShift} we see that the possible
centroids measured by the different observers fill a disk of radius
$R_{max}=S_{\star}/M$ about the point $x_{CM}^{\alpha}(P)$; every
point of such disk could also be a proper center of mass, provided
that it rotates with angular velocity $\vec{\omega}$. This is the
result found by Möller~\cite{MollerAIP}. In a frame where $P^{i}\ne0$
(i.e., moving relative to $\mathcal{O}$) this leads to helical motions,
as depicted in Fig.~\ref{fig:MolerPicture}, which are precisely
the ones explicitly derived in the previous section. We emphasize
that the angular velocity $\vec{\omega}$ of the disk of proper centers
of mass is \emph{not} the same as the angular velocity the body; indeed
it is \emph{opposite} to the sense of rotation of the body; and note
also that the points of the circle $R_{max}=S_{\star}/M$ move at
the speed of light.

Finally it is clear, from the analysis above, that \emph{all the helical
solutions are contained within a tube of radius} $R_{max}=S_{\star}/M$,
which is actually (see~\cite{MollerBook}) the \emph{minimum} size
a classical spinning particle can have if it is to have finite $S_{\star}$
and positive mass without violating the laws of Special Relativity.

\section{Dynamical interpretation of the helical motions\label{sec:Dynamical-interpretation-of}}

%
\begin{figure}
\includegraphics[width=0.45\textwidth]{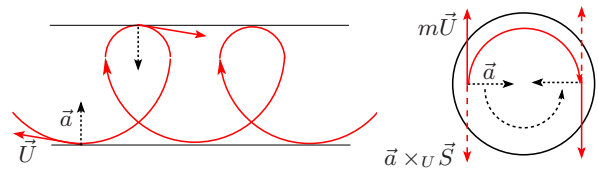} 

\caption{\label{fig:HiddenHelical}Hidden {}``inertial'' momentum provides
dynamical interpretation for the helical motions (left panel): the
acceleration results from an interchange between kinetic $P_{kin}^{\alpha}=mU^{\alpha}$
and (inertial) hidden momentum $P_{hid}^{\alpha}=S^{\alpha\beta}a_{\beta}$,
which occurs in a way that their variations cancel out at every instant,
keeping the total momentum constant. This is made manifest in the
right panel, representing the $\vec{P}=0$ frame, wherein $\vec{P}_{hid}=\vec{a}\times_{U}\!\vec{S}=-m\vec{U}=-\vec{P}_{kin}$.
The description is formally analogous to the bobbing~\cite{Wald et al 2010}
of a magnetic dipole orbiting a cylindrical charge. }

\end{figure}

The concept of hidden momentum is central to the understanding of
dynamics of the helical solutions; namely it explains how the motion
of a free spinning particle can be consistently described by helical
solutions without violating any conservation principle, and that they
are a phenomena which can be cast as analogous to the bobbing of a
magnetic dipole in an external electric field studied in~\cite{Wald et al 2010}.
As we have seen in the previous sections, for a \emph{free} spinning
particle in flat spacetime, Eqs.~\eqref{eq:Eqs Motion Flat}, supplemented
with $S^{\alpha\beta}U_{\beta}=0$, yield, as possible solutions,
straight line motion plus a set of helical motions contained within
a tube of radius $S_{\star}/M$. This seems odd at first glance: how
can a solution where the center of mass of the particle is accelerating
without any external force be physically acceptable? The answer is
that the acceleration results from an interchange between kinetic
momentum $mU^{\alpha}$ and hidden momentum $S^{\alpha\beta}a_{\beta}$
(we dub it {}``inertial'' hidden momentum, the reason for such denomination
being explained below) which occurs in a way such that their variations
cancel out at every instant, keeping the total momentum constant,
as illustrated in Fig.~\ref{fig:HiddenHelical}. That is what we
are going to show next. Consider a generic spin condition $S^{\alpha\beta}u_{\beta}=0$,
where $u^{\alpha}$ denotes the 4-velocity of an arbitrary observer
$\mathcal{O}(u)$. As discussed in Sec.~\ref{sub:Center-of-mass},
this condition means that we take as reference worldline the center
of mass as measured by $\mathcal{O}(u)$. Contracting (\ref{eq:Eqs Motion Flat}b)
with $u_{\beta}$, and using $S^{\alpha\beta}u_{\beta}=0\Rightarrow u_{\beta}DS^{\alpha\beta}/d\tau=-S^{\alpha\beta}Du_{\beta}/d\tau$,
leads to \begin{equation}
\ S^{\alpha\beta}\frac{Du_{\beta}}{d\tau}=\gamma(u,U)P^{\alpha}-m(u)U^{\alpha}\label{eq:HiddenInertial}\end{equation}
 where $\gamma(u,U)\equiv-U^{\beta}u_{\beta}$ and $m(u)\equiv-P^{\beta}u_{\beta}$.
Hence, if $Du_{\beta}/d\tau\ne0$, then in general the momentum is
not parallel to the 4-velocity: $P^{[\alpha}U^{\beta]}\ne0$, and
the particle is said to have \emph{hidden momentum}. The momentum
of the particle may be split in its projections parallel and orthogonal
to the CM 4-velocity $U^{\alpha}$: \begin{equation}
P^{\alpha}=P_{kin}^{\alpha}+P_{hid}^{\alpha};\quad P_{kin}^{\alpha}\equiv mU^{\alpha},\; P_{hid}^{\alpha}\equiv(h^{U})_{\ \beta}^{\alpha}P^{\beta}.\label{eq:HiddenMomentum}\end{equation}
 where $\ (h^{U})_{\ \beta}^{\alpha}\equiv U^{\alpha}U_{\beta}+\delta_{\ \beta}^{\alpha}$
denotes the space projector with respect to $U^{\alpha}$. We dub
the time projection $P_{kin}^{\alpha}=mU^{\alpha}$ {}``kinetic momentum''
associated with the motion of the centroid; and the component $P_{hid}^{\alpha}$
orthogonal to $U^{\alpha}$ is what we dub {}``hidden momentum''.
The reason for the latter denomination is easily seen taking the perspective
of an observer $\mathcal{O}(U)$ comoving with the particle: in the
frame of $\mathcal{O}(U)$ (i.e., the frame $U^{i}=0$) the 3-momentum
is in general not zero: $\vec{P}=\vec{P}_{hid}\ne0$; however, by
definition, the particle's CM is at rest in that frame; such momentum
must thus be hidden somehow.

Now, if $Du_{\beta}/d\tau=0$, that is, if we take as reference worldline
the center of mass as measured by an observer $\mathcal{O}(u)$ \emph{such
that} $u^{\alpha}$ \emph{is parallel transported along it}, then
from Eq. (\ref{eq:HiddenInertial}) we have $P^{\alpha}\parallel U^{\alpha}$,
and there is no hidden momentum. This is actually cast in~\cite{Semerak II}
as one of the possible spin supplementary conditions. Thus indeed
this form of hidden momentum is pure gauge; it also means that the
motion effects induced by it (such as the bobbings studied in~\cite{Wald et al 2010})
must be confined to the worldtube of centroids, so that they can be
made to vanish by a suitable choice of reference worldline. For this
reason it is dubbed in~\cite{Wald et al 2010} {}``kinematical hidden
momentum'' (by contrast with gauge invariant hidden momentum present
in electromagnetic systems, dubbed {}``dynamic'' therein). In flat
spacetime, we can say that if the observer is inertial (which implies
that its 4-velocity is parallel transported along the particle's worldline),
then there is no hidden momentum. (The {}``laboratory'' observer
considered in p. 9 of~\cite{Wald et al 2010}, for the case of flat
spacetime, is an example of an inertial observer, more precisely the
static observer with $u^{\alpha}$ tangent to the time Killing vector).

This hidden momentum is, of course, related with the relativity of
the center of mass (its shift in different frames, discussed in Sec.~\ref{sub:Center-of-mass}),
and taking this perspective makes quite clear why $U^{\alpha}$ decouples
from $P^{\alpha}$ if $Du^{\alpha}/d\tau\ne0$, and the hidden momentum
arises. But first let us make some remarks:

\emph{Remark.}--- Whereas in the previous sections we dealt essentially
with flat spacetime, Eq.~\eqref{eq:HiddenInertial} above is general.
In the previous sections we illustrated the non-uniqueness of the
center of mass by its relativity with respect to what we called {}``observers''
and/or {}``frames'', and that their {}``acceleration'' was the
underlying reason behind the non-trivial velocity the centroid has
in some cases. But what we are implicitly doing (and what actually
holds in a more general formulation e.g.~\cite{Dixon1964,Dixon1970I,Wald et al 2010,Gyros}),
is to assume a continuous field of time-like unit vectors $u^{\alpha}$
along $z^{\alpha}(\tau)$; at each event, $u^{\alpha}$ provides the
hypersurface $\Sigma(u,\tau)$ over which the integrals defining moments
$P^{\alpha},\ S^{\alpha\beta}$ (as well as the center of mass) are
performed. $\Sigma(u,\tau)$ is generically defined as the hypersurface
formed by all geodesics orthogonal to $u^{\alpha}$ at the point $z^{\alpha}(\tau)$.
Thus in this construction, the vectors $u^{\alpha}$ (which we can
always think about as the instantaneous 4-velocity of some \emph{local}
observer) are all that matter; the concept of an observer $\mathcal{O}(u)$,
in the traditional sense of a worldline to which $u^{\alpha}$ is
tangent, has no place; except for the case $u^{\alpha}=U^{\alpha}\equiv dz^{\alpha}/d\tau$,
there is no worldline tangent to the field $u^{\alpha}$ (and therefore
no acceleration is defined for it). The field $u^{\alpha}$ only has
to exist along the reference worldline, and $Du^{\alpha}/d\tau$ is
the only derivative defined for it. In the special case of flat spacetime
(but not in general curved spacetime!), where vectors at different
points can be compared, we can indeed think of the field $u^{\alpha}$
as the tangent to the worldline of some distant (as such worldline
in general will not coincide with $z^{\alpha}$) observer, and $Dv^{\alpha}/d\tau$
as its coordinate acceleration with respect to the CM frame $U^{i}=0$.
This is what was implicitly done in Sec. \ref{sec:Kinematical-explanation-of}.

As we have seen in Sec.~\ref{sub:Center-of-mass}, the position of
the centroid of a spinning body depends on the vector $u^{\alpha}$
relative to which it is computed. If that vector varies along the
reference worldline it is clear that this is reflected in the velocity
$U^{\alpha}$ of the centroid, which in general will accelerate even
without the action of any force. Also $U^{\alpha}$ will in general
no longer be parallel to $P^{\alpha}$, i.e., $\vec{U}\ne0$, and
thus the centroid is not at rest in the frame $P^{i}=0$. Let us show
explicitly that the decoupling of $U^{\alpha}$ from $P^{\alpha}$,
manifest in Eq.~\eqref{eq:HiddenInertial}, indeed comes from the
shift of the centroid, given in Eq.~\eqref{eq:CMCovariantShift}.
As we have seen in Sec. \ref{sub:Center-of-mass}, if we choose the
reference line to be the center of momentum centroid, $z^{\alpha}=x_{CM}^{\alpha}(P)$,
then the shift of the centroid measured by the observer $u^{\alpha}$
is $\Delta x^{\alpha}=-S_{\star}^{\alpha\beta}u_{\beta}/m(u)$, with
$m(u)\equiv-P^{\alpha}u_{\alpha}$. Therefore: \begin{equation}
\frac{D\Delta x^{\alpha}}{d\tau}=-\frac{S_{\star}^{\alpha\beta}}{m(u)}\frac{Du_{\beta}}{d\tau}+\frac{dm(u)}{d\tau}\frac{S_{\star}^{\alpha\beta}u_{\beta}}{m(u)^{2}},\label{eq:DerivativeShift}\end{equation}
 as by \eqref{eq:Eqs Motion Flat}, $DS_{\star}^{\alpha\beta}/d\tau=0$.
By \eqref{eq:SpinTensorRelation} we know that the spin tensor $S^{\alpha\beta}$
computed by $u^{\alpha}$ satisfies \[
S^{\alpha\beta}=S_{\star}^{\alpha\beta}-\Delta x^{\alpha}P^{\beta}+P^{\alpha}\Delta x^{\beta}.\]
 Substituting in \eqref{eq:DerivativeShift}, using $S^{\alpha\beta}u_{\beta}=0$
and $\Delta x^{\beta}u_{\beta}=0$, we obtain \begin{eqnarray*}
m(u)\frac{D\Delta x^{\alpha}}{d\tau} & = & -S^{\alpha\beta}\frac{Du_{\beta}}{d\tau}-\Delta x^{\alpha}P^{\beta}\frac{Du_{\beta}}{d\tau}\\
 &  & -P^{\alpha}\frac{D\Delta x^{\beta}}{d\tau}u_{\beta}-\frac{dm(u)}{d\tau}\Delta x^{\alpha},\end{eqnarray*}
 or, noticing that $dm(u)/d\tau=-P^{\beta}Du_{\beta}/d\tau$, we have
\begin{equation}
m(u)\frac{D\Delta x^{\alpha}}{d\tau}=-S^{\alpha\beta}\frac{Du_{\beta}}{d\tau}-P^{\alpha}\frac{D\Delta x^{\beta}}{d\tau}u_{\beta}.\label{eq:DerivativeShift2}\end{equation}
 In flat spacetime and Cartesian coordinates, we may always write:\begin{equation}
U^{\alpha}=\frac{dx_{CM}^{\alpha}(u)}{d\tau}=\frac{dx_{CM}^{\alpha}(P)}{d\tau}+\frac{d\Delta x^{\alpha}}{d\tau}=fP^{\alpha}+\frac{D\Delta x^{\alpha}}{d\tau}\label{eq:f}\end{equation}
 where $f$ is a function to be determined. Since $\Delta x^{\alpha}P_{\alpha}=0$
and $DP^{\alpha}/d\tau=0$, it follows that $P_{\alpha}D\Delta x^{\alpha}/d\tau=0$;
thus contracting (\ref{eq:f}) with $P_{\alpha}$ we obtain $f=m/M^{2}$.
Finally, substituting Eq. (\ref{eq:f}) for $D\Delta x^{\alpha}/d\tau$
in \eqref{eq:DerivativeShift2}, we obtain \eqref{eq:HiddenInertial}
exactly.

Hence we have different, and equivalent, descriptions for the same
motion (of a free particle in flat spacetime). The most simple ones
are the centroids measured by every possible inertial observers, whose
trajectories are straight lines parallel to each other, and to $P^{\alpha}$.
In the frame $P^{i}=0,$ all these centroids are at rest. But if we
take the centroid with respect to an $u^{\alpha}$ not constant along
the curve, which, as discussed above, may be thought as the point
of view of some accelerated observer $\mathcal{O}(u)$, then the centroid
will have in general a different velocity, and also accelerate, see
Eqs. (\ref{eq:ShiftVel}). However $P^{\alpha}$ is always the same
(it does not depend on the choice of centroid)! This makes evident
the role of $P_{hid}^{\alpha}$ in a consistent dynamical description:
when one describes the body through the centroid measured by an accelerated
observer, \emph{there must be a hidden momentum} $\vec{P}_{hid}$
that cancels out the kinetic momentum $\vec{P}_{kin}=m\vec{U}$ the
moving centroid $x_{CM}^{\alpha}(u)$ has in the frame $P^{i}=0$.

If the observer's acceleration itself changes in a way such that the
signal in Eq. (\ref{eq:ShiftVel}) oscillates, we may have a bobbing;
or if it is such that $\mathcal{O}(u)$ sees its centroid to be at
rest (i.e., if $\mathcal{O}(u)$ moves with 3-velocity (\ref{eq:VelHelical})
in the frame $P^{i}=0$), then we have a helical solution. In this
case ($u^{\alpha}=U^{\alpha}$), decomposition (\ref{eq:HiddenMomentum})
takes a simple form, cf. Eq. (\ref{eq:momentumhelical}): \begin{equation}
P_{kin}^{\alpha}=mU^{\alpha},\quad P_{hid}^{\alpha}=S^{\alpha\beta}a_{\beta}=\epsilon_{\ \ \gamma\delta}^{\alpha\beta}a_{\beta}S^{\gamma}U^{\delta}\ ;\label{eq:PhidPirani}\end{equation}
 which in the frame $U^{i}=0$ reads $\vec{P}=\vec{P}_{hid}=\vec{a}\times\vec{S}$.
Since $\vec{G}=-\vec{a}$ is the {}``gravitoelectric'' field~\cite{The many faces,Natario}
as measured in that frame (which is a field of {}``inertial'' forces),
$\vec{P}_{hid}$ is cast in~\cite{Gyros} as the inertial analogue
of the hidden momentum $\vec{\mu}\times\vec{E}$ of electromagnetic
systems (see e.g.~\cite{Vaidman,Wald et al 2010}), and its origin
explained therein by an analogous model. In this spirit, the dynamics
of the helical representations may be cast as analogous to the bobbing
of a magnetic dipole orbiting a cylindrical charge, discussed in Sec.~IIIB.1
of~\cite{Wald et al 2010}. Let the line charge be along the $z$
axis, and $\vec{E}$ the electric field it produces; and consider
an oppositely charged test particle, with magnetic dipole moment $\vec{\mu}=(\mu^{x},\mu^{y},0)$,
orbiting it. The $z$ component of the force vanishes for this setup;
hence $P^{z}=0=constant$. But the particle will possess a hidden
momentum, which for slow motion~\cite{Wald et al 2010} reads $\vec{P}_{hid}=\vec{\mu}\times\vec{E}$;
as it orbits the line charge, $\vec{P}_{hid}$ oscillates between
positive and negative values along the $z$-axis, implying the particle
to bob up and down in order to keep the total momentum along $z$
constant: $P^{z}=P_{kin}^{z}+P_{hid}^{z}=0$. Thus, just like in the
case of the helical motions, the bobbing arises not through the action
of a force, but from an interchange between kinetic $\vec{P}_{kin}=m\vec{U}$
and hidden momentum. The difference being that the hidden momentum
$\vec{S}\times\vec{a}$ is \emph{pure gauge} (which indeed allows
a helical solution to be a consistent description of the motion even
in the case of a free particle in flat spacetime, but can always be
made to vanish by choosing the non-helical representation), whereas,
by contrast, the electromagnetic effect mentioned above is physical
and \emph{gauge independent}.

Hence again we see that the straight line and helical solutions are
\emph{alternative} and \emph{physically consistent} descriptions of
the motion of a free spinning body: in the first case, we have no
acceleration and no hidden momentum; in the second case we have an
helix, but also inertial hidden momentum.

\section{Conclusion. Misconceptions about the helical solutions and quantum
zitterbewegung}

Mathisson's helical solutions have been deemed unphysical in number
of treatments, e.g.~\cite{Dixon1964,Dixon1965}, due to the wrong
idea that Eq.~\eqref{eq:RadiusWeyssenhoff}, which is equivalent
to Eq.~\eqref{eq:Radius Fixed}, allows the helical motions of (the
representative point of) a given particle to have an arbitrarily large
radius~\cite{WeyssenhoffNature,WeyssenhoffRaabe,Dixon1964,Dixon1965,DixonReview}.
This would also imply that these solutions were not equivalent to
the ones derived in Sec.~\ref{sec:Kinematical-explanation-of}, and
found in Möller's treatment~\cite{MollerAIP}. As we have seen, this
is just a misconception, based on the failure to notice that in order
to have a set of helical solutions representing the same physical
body, we must fix $\gamma S=S_{\star}$ and $m/\gamma=M$, not $m$
and $S$; i.e., we must require that, regardless of the different
possible representations, it has the \emph{same moments} as measured
with respect to the \emph{same observer} and \emph{reference worldline}.

There is nothing unphysical with Mathisson's helical solutions; they
are all perfectly valid and equivalent descriptions of the motion
of a classical spinning body. The helices are, as we have seen, all
contained within a worldtube of radius $R_{max}=S_{\star}/M$ centered
at $x^{\alpha}(P)$ (i.e., the center of mass as measured in the zero
3-momentum frame). And that should be a natural result from the analysis
in Sec.~\ref{sub:Center-of-mass}: the radius of a helical motion
of 4-velocity $U^{\alpha}$ corresponds to the displacement of the
center of mass measured in the frame $U^{i}=0$ relative to $x^{\alpha}(P)$;
the maximum shift is $\Delta x_{max}=S_{\star}/M$, corresponding
to the case that the relative velocity between the two observers is
the speed of light. Now one also has to note that $R_{max}=S_{\star}/M$
is also (see~\cite{MollerBook,MollerAIP}) the \emph{minimum} size
a that classical spinning particle can have if it is to have finite
$S_{\star}$ and positive mass without violating Special Relativity
(i.e., without containing points moving faster than the speed of light).
This means that not only $R$ is not arbitrarily large, but also it
can never exceed the minimum size of the particle. i.e., the helical
trajectories always fall \emph{within} the body. Furthermore, they
have a clear kinematical explanation as shown in Sec.~\ref{sec:Kinematical-explanation-of};
and their dynamics may be interpreted in analogy with the hidden momentum
of electromagnetic systems~\cite{Gyros}.

The helical solutions were interpreted by some authors~\cite{Mathisson Zitterbewegung,Corben,CorbenBook,CorbenQRotor,Papapetrou I}
as the classical limit of the quantum zitterbewegung, due to the similarity
between the zitterbewegung frequency of the Dirac equation for the
electron and the frequency of the corresponding classical helical
motions. Indeed, putting $S_{\star}=\hbar/2$ and $M=M_{e}$, we obtain
$\omega=2M_{e}/\hbar$, which is precisely Dirac's zitterbewegung
frequency for the electron (this extends Mathisson's observation in
\cite{Mathisson Zitterbewegung} to the relativistic limit). Other
authors~\cite{WeyssenhoffNature,WeyssenhoffRaabe,Dixon1964,Dixon1965}
have rejected this correspondence, based on two arguments: 1) that
Mathisson's helical solutions for the electron might have arbitrarily
large radius which would make them macroscopically measurable~\cite{WeyssenhoffNature,WeyssenhoffRaabe,Dixon1964,Dixon1965};
2) that the coincidence between the frequencies holds only in the
non-relativistic limit~\cite{WeyssenhoffNature,WeyssenhoffRaabe}
(based on the expression (\ref{eq:wMathisson}) in the form $\omega=m/\gamma^{2}S$,
and repeating the same misunderstanding that led to the arbitrary
radius). A deeper analysis of this problem will be presented elsewhere.
Herein we would just like to point out that, as made clear by the
analysis above, both arguments put forward against this correspondence
between classical and quantum solutions arise from misconceptions.

Finally, an aspect that has drawn skepticism (see e.g.~\cite{Wald et al,Wald et al 2010}{})
into the equations of motion supplemented with the Mathisson-Pirani
condition is the fact that they are of third order, meaning that,
in order for the motion to be determined, not only one must prescribe
the initial position and velocity, but also (in general) the initial
acceleration. This might seem odd as in Newtonian mechanics (where
the center of mass is an invariant) the motion of the CM of an extended
body is fully determined by the force laws given its initial position
and velocity. But in Relativity that is only true for a monopole particle;
for a general extended (spinning) body, due to the relativity of the
center of mass, in addition to those two initial conditions, one needs
also to determine the field of unit time-like vectors $u^{\alpha}$
relative to which the CM is computed. It is important to note that,
as explained in Sec. \ref{sec:Dynamical-interpretation-of}, the acceleration
of the CM does not originate solely from the force, but also from
the variation of the field $u^{\alpha}$ along the CM worldline (leading
to the hidden momentum). Some spin conditions, such as the Tulczyjew-Dixon
\cite{Tulczyjew,Dixon1964,Dixon1970I} or the Corinaldesi-Papapetrou
conditions~\cite{Corinaldesi Papapetrou}, fully fix the reference
worldline and the vector field $u^{\alpha}$ along it; the Mathisson-Pirani
condition, as explained in Sec. \ref{sec:Kinematical-explanation-of},
does not, and the higher order of the equations merely reflects that
incompleteness of gauge fixing.

\emph{Why does it matter? ---} In addiction to the physical clarification
of the helical motions, an important point made in this work is to
prove the physical validity of the Mathisson-Pirani condition. We
have shown it is as valid as any other of the infinite number of possible
spin conditions. Indeed, a condition of the type $S^{\alpha\beta}u_{\beta}=0$,
for some unit time-like vector $u^{\alpha}$, amounts to choosing
as the representative point of the body the center of mass as measured
by some observer of 4-velocity $u^{\alpha}$. But $u^{\alpha}$ is
arbitrary; whether it is itself the 4-velocity of the center of mass
$U^{\alpha}$ (Mathisson-Pirani condition), or it is parallel to $P^{\alpha}$
(Tulczyjew-Dixon condition~\cite{Tulczyjew,Dixon1964,Dixon1970I}),
or it corresponds to the static observers in Schwarzschild spacetime
(Papapetrou-Corinaldesi condition~\cite{Corinaldesi Papapetrou}),
or any other type of observers (there is an infinite number of possibilities),
the choice should be based on convenience%
\footnote{In this work we dealt with free particles in flat spacetime, where
it was clear that all the centroids (including those corresponding
to the helical motions) remain inside the worldtube of radius $S_{\star}/M$
forever. However in the presence of strongly inhomogeneous external
fields, the point we chose to represent the particle makes a difference.
The trajectories are seen to diverge (outside any such worldtube)
in~\cite{Semerak II} for a Kerr background. This actually signals
the breakdown of the pole-dipole approximations (not that the spin
conditions are not pure gauge after all!). The approximation is only
acceptable when the choice of centroid (and the spin condition) does
not matter; i.e., when the scale of variation of the external field
is much larger than $S_{\star}/M$, cf.~\cite{Semerak II}.%
}. So the question might be posed, why worry about Mathisson-Pirani's
condition, which leads to degenerated solutions, while the Tulczyjew-Dixon
condition yields a unique definition of CM? The point is that there
are also situations where it is the Mathisson-Pirani condition that
gives the simplest and more enlightening solution. For a free particle
in flat spacetime, indeed the Tulczyjew-Dixon condition $S^{\alpha\beta}P_{\beta}=0$
provides the simplest description for the center of mass motion, which
is uniform straight line motion (and coincides \emph{in this special
case} with Mathisson's non-helical solution), whereas the Mathisson-Pirani
condition includes also the helical solutions, which are more complicated
descriptions. However in the presence of gravitational and electromagnetic
fields, the Tulczyjew-Dixon solution no longer coincides with any
of Mathisson's solutions%
\footnote{When a electromagnetic and/or gravitational field (or any other external
force) are present, $x_{CM}^{\alpha}(P)$ is \emph{not,} in general,
a proper center of mass; i.e., the 4-velocity of the centroid defined
by $S^{\alpha\beta}P_{\beta}=0$ is not parallel to $P^{\alpha}$;
that can be seen from e.g. Eq.~{(35)} of~\cite{Wald et al 2010}
(see also discussion in~\cite{Gyros}). In other words, the centroid
measured in the $P^{i}=0$ frame is not at rest in that frame.%
}, and it turns out, as exemplified in several applications in~\cite{Gyros},
that in some more complex setups it is the Mathisson-Pirani condition
that provides the simplest and clearest description. This condition
arises also in a natural fashion in a number of treatments~\cite{Taub,EulerTop}
(see also~\cite{Plyatsko Non-Oscillatory}); for massless particles,
it has been argued in~\cite{BaylinMassless,BaylinMassless II} that
it is actually the only one that can be applied. And for the case
of the equation for the spin evolution (\ref{eq:Eqs Motion Flat}b),
it is \emph{always} the Mathisson-Pirani condition that yields the
simplest and physically more sound description: in the absence of
electromagnetic field (or other external torques), $S^{\alpha}$ is
Fermi-Walker transported; i.e., the gyroscope's axis is fixed relative
to a non-rotating frame, which is the natural, expected result. The
Tulczyjew-Dixon condition yields a different equation (see Eq.~{(7.11)}
of~\cite{Dixon1970I}), meaning that $S^{\alpha}$ undergoes transport
orthogonal to $P^{\alpha}$ (dubbed therein {}``M-transport'').
There is no conflict because the transport is along a different worldline.
But the equation for the evolution of $S^{\alpha}$ puts a strong
emphasis on the relevance of acknowledging the physical validity of
the Mathisson-Pirani condition, which has to do has to do with the
deepest notions of inertia and rotation in General Relativity: a Fermi-Walker
transported frame is by definition a frame that does not rotate relative
to the local spacetime (i.e., to the {}``local compass of inertia'',
as described in some literature, e.g.~\cite{Gravitation and Inertia});
in order for this law to be more than a mere mathematical abstraction,
and for the rotation to be something absolute and locally measurable,
this has to have a correspondence to a physical object, which is the
gyroscope (gyroscopes are objects that oppose to changes in direction
of their axis of rotation!). \emph{Only} if the Mathisson-Pirani condition
holds is a gyroscope Fermi-Walker transported. Hence, deeming the
Mathisson-Pirani condition as physically unacceptable (as many authors
do), amounts to saying that the whole concept of Fermi-Walker transport
makes no sense from the physical point of view (or is at best an approximation).

Finally, probably one of the most interesting features of the Mathisson-Pirani
condition is the fact that it makes explicit two \emph{exact} gravito-electromagnetic
analogies: the tidal tensor analogy relating the gravitational force
on a spinning particle with the electromagnetic force exerted on a
magnetic dipole, discussed in~\cite{OurPRD,Gyros}, and the analogy
based on the 1+3 formalism discussed in e.g.~\cite{Natario,The many faces,Gyros},
the latter with the following realizations: one again relating the
two forces~\cite{Natario}, other relating the evolution of the spin
of a gyroscope with the precession of a magnetic dipole under the
action of a magnetic field~\cite{Natario,The many faces,Gyros},
and a third one relating the hidden inertial momentum with the hidden
momentum of electromagnetic systems~\cite{Gyros}. These analogies
provide valuable insight, and a familiar formalism to treat otherwise
exotic gravitational effects, as well as a means to contrast them
with their electromagnetic counterparts. Such comparison allows one
to notice some fundamental aspects of both interactions, as explained
in detail in~\cite{Gyros}.

\section*{Acknowledgments}

We thank João Penedones for discussions. This work was partially supported
by FCT-Portugal through project CERN/FP/116341/2010. L.F.C.\ and
M.Z.\ are funded by FCT through grants SFRH/BD/41370/2007 and SFRH/BD/43558/2008.

\end{document}